\documentstyle[amsfonts,amsmath,amsfonts,amssymb]{article} 

\begin{document}
\title{Thermodynamic Bethe ansatz equation for $osp(1|2)$ 
integrable spin chain}
\author{Kazumitsu Sakai
\footnote{
e-mail address: sakai@as2.c.u-tokyo.ac.jp}  
 and Zengo Tsuboi\footnote{
JSPS Research Fellow; 
e-mail address: tsuboi@gokutan.c.u-tokyo.ac.jp} \\
Institute of Physics,                                   
 University of Tokyo, \\
 Komaba   3-8-1, Meguro-ku, Tokyo 153-8902, Japan}
\date{}
\maketitle
\begin{abstract}
The thermodynamic Bethe ansatz is applied to a 
quantum integrable spin chain associated with 
 the Lie superalgebra $osp(1|2)$. 
Using the string hypothesis, we derive a set of 
infinite number of  non-linear integral equations 
(thermodynamic Bethe ansatz equation), which 
characterize the free energy.
The low temperature limit of the free energy 
is also discussed.
\end{abstract} 
Modern Physics Letters A, in press.

\vspace{20pt}

\noindent 
Solvable lattice models related to Lie superalgebras \cite{Ka} 
have  received much attentions
\cite{KulSk,Kul,BS,KuR,DFI,Sa,ZBG,MR94}. 
Characteristically, these models possess both fermionic 
and bosonic degree of freedom and their Boltzmann weights 
($R$-matrices) satisfy the 
 graded Yang-Baxter equations\cite{KulSk}. 
Some of them include  well-known models 
as special cases.  
For example, supersymmetric $t-J$ model is related 
to the Lie superalgebra $sl(1|2)$. 
To analyze such models, Bethe ansatz is often used 
(See for example \cite{Kul,Sch,EK,EKS,FK,EKtba,Ma,Mar95-1,Mar95-2,RM,PF,MR,T1,T2,T3,T4,T5,JKS,MNR,Sa99} and references therein).  
Moreover, thermodynamics  
have been discussed by several people.  
In particular, the thermodynamic Bethe ansatz 
 (TBA) equation was given for the supersymmetric $t-J$ model 
 \cite{Sch,JKS} and the supersymmetric extended 
 Hubbard model \cite{EKtba,JKS}, 
 which is related to $sl(2|2)$. 
 More recently, the TBA equation for $sl(r|s)$ model
  was discussed \cite{Sa99} in 
 relation with the continuum limit of integrable spin chains. 
 
 However, in contrast to $sl(r|s)$ case, 
 the thermodynamics of quantum spin chain related to 
 the orthosymplectic Lie superalgebra $osp(r|2s)$ 
 is not much understood. 
 In particular, to the author's knowledge, 
 there is no literature on the thermodynamic Bethe ansatz 
 (TBA) equation even for the simplest orthosymplectic 
 $osp(1|2)$ integrable spin chain 
 \cite{Kul,BS,KuR,Sa,ZBG,MR94,Mar95-1,Mar95-2}. 
 This is regrettable because this model may 
 be related to interesting problems such as 
 $N=1$ supersymmetry in particle physics, and 
 the loop model, which will describe 
 statistical properties of polymers 
 in condensed matter physics \cite{MNR}. 

 The purpose of this letter is to 
  construct the 
 TBA equation for $osp(1|2)$ integrable spin chain  
 based on the string hypothesis\cite{T,G,TS}. 
  The resultant TBA equation, which describes the free energy 
  is a set of 
  infinite number of non-linear integral equations. 
  We have also confirmed the fact that our TBA equation 
coincides with the one \cite{IP} 
from the quantum transfer matrix 
 (QTM) approach \cite{Klu,JKS,KSS,sakai} 
 based on the transfer matrix 
 functional relations \cite{T5,T1,T2,T3,T4} 
 ($osp(1|2)$ version of the $T$-system). 
This letter yields a  basis of future studies 
 of thermodynamics for more general models such as 
 $osp(r|2s)$ model. 

Now we shall describe the $osp(1|2)$ 
 model\cite{Kul,BS,KuR,Sa,ZBG,MR94,Mar95-1,Mar95-2}. 
The $\check{R}$-matrix of this model is given as 
a rational solution of the graded Yang-Baxter 
equation \cite{KulSk} 
associated with the 
three dimensional representation of the Lie superalgebra 
$osp(1|2)$ (to be precise, \symbol{96}super Yangian' $Y(osp(1|2))$), 
whose basis is ${\mathbb Z}_{2}$-graded and 
labeled by the parameters $p(1)=p(3)=1,p(2)=0$. 
One can construct the $\check{R}(u)$-matrix as a
 \symbol{96}Baxterization' of the Temperley-Lieb generator
 \cite{MR94}. 
It reads explicitly as follows:
\begin{eqnarray}
\check{R}(u)=I+uP^{g}-\frac{u}{u-\frac{3}{2}}E^{g}, \label{rmat}
\end{eqnarray}
where  $P^{g}$ denotes the graded permutation 
operator which has $9\times 9$ matrix elements 
$ (P^{g})_{ab}^{cd}=(-1)^{p(a)p(b)} \delta_{a,d}\delta_{b,c}$;  
 $E^{g}$ is a $9\times 9$ matrix whose matrix elements are 
$ (E^{g})_{ab}^{cd}=\alpha_{ab} \alpha_{cd}^{-1}$; 
 $\alpha $ is a $3 \times 3$ matrix:
\begin{eqnarray}
\alpha=
\left(
 \begin{array}{ccc}
  0  & 0 & 1 \\ 
  0  & 1 & 0 \\ 
  -1 & 0 & 0 
 \end{array}
\right). \label{jirou}
\end{eqnarray}
There are two kinds of the definition of the transfer matrix 
for each model related to Lie superalgebras.  
The one is defined as the supertrace of a monodromy matrix 
and the other is the ordinary trace of a monodromy matrix. 
In this letter, we adopt the latter. 
In this case, the transfer matrix $T(u)$ of the corresponding system 
can be written as 
\begin{equation}
T(u)=\mbox{tr}_a [R_{aN}(iu)\cdots R_{a 1}(iu)], \label{transf}
\end{equation}
where $N$ is the number of lattice sites; 
$a$ denotes the auxiliary space; $R_{aj}(iu)$ 
denotes $R(iu)$ which acts non-trivially on 
the auxiliary space and $j$-th site of the quantum space; 
 $R(u)=P \check{R}(u)$: 
 $P_{ab}^{cd}=\delta_{a,d}\delta_{b,c}$ is the (non-graded) 
permutation operator. 

The Hamiltonian is defined by 
taking the logarithmic derivative of above transfer matrix at $u=0$,
\begin{eqnarray}
H=\frac{J}{i}\frac{d}{du}\ln T(u) \biggl|_{u=0}
 =J\sum_{j=1}^{N}
 \left(
   P^{g}_{j, j+1}+\frac{2}{3}E^{g}_{j, j+1}
 \right) \label{hami}, 
\end{eqnarray}
where we assume a periodic boundary condition.
 Here $J$ is a real coupling constant 
which characterizes the phase of the model;
the ferromagnetic and antiferromagnetic regimes correspond 
$J>0$ and $J<0$, respectively. 
The eigenvalue formula of (\ref{transf}) 
 (also written as $T(u)$)
is given \cite{Kul,Mar95-2,MR} in the dressed vacuum form 
 as follows 
\begin{eqnarray}
T(u)&=&(-1)^{N-n}(u+i)^{N}
\frac{Q(u-\frac{i}{2})}
     {Q( u+\frac{i}{2})}
+u^{N}
\frac{Q(u)Q(u+\frac{3i}{2})}
     {Q(u+\frac{i}{2})Q(u+i)}
     \nonumber \\ 
&& 
+(-1)^{N-n}
\left(\frac{u(u+\frac{i}{2})}{u+\frac{3i}{2}}\right)^{N}
\frac{Q(u+2i)}
     {Q(u+i)},
     \label{DVF}
\end{eqnarray}
where $ Q(u)= \prod_{j=1}^{n}(u-u_{j})$; 
 $n \in \{0,1,\dots, N \}$ 
 is a quantum number; $u_{j} \in {\mathbb C}$.  
The dressed vacuum form is built on the pseudo-vacuum state 
corresponding $n=0$. 
This formula (\ref{DVF}) 
is free of poles under the following Bethe ansatz 
equation (BAE) 
\begin{eqnarray}
\left( \frac{u_{k}+\frac{i}{2}}{u_{k}-\frac{i}{2}} \right)^{N}
 = 
 -(-1)^{N-n}
\frac{Q(u_{k}-\frac{i}{2})Q(u_{k}+i)}
     {Q(u_{k}+\frac{i}{2})Q(u_{k}-i)} \quad 
     {\rm for } \quad k \in \{1,2,\dots,n\} 
       \label{BAE}.
\end{eqnarray}
Taking the logarithmic derivative of (\ref{DVF}), 
we obtain the 
eigenvalue formula of the Hamiltonian (\ref{hami})
\begin{eqnarray}
\hspace{-40pt}
&& E=\frac{J}{i} \frac{d}{du} \ln T(u) \Big|_{u=0}=
 J \left\{
 \sum_{j=1}^{n}
 \frac{1}{(u_{j})^{2}+\frac{1}{4}}
 -N
 \right\} . 
 \label{en} 
\end{eqnarray} 
Following Takahashi's \cite{T} argument 
(see also \cite{YY}), 
we shall discuss the thermodynamics 
  based on the string hypothesis.  
The $m$-string solution of the BAE (\ref{BAE}) is
 given \cite{Mar95-1,Mar95-2} 
as follows 
\begin{eqnarray}
u_{k}^{m,\alpha}=u_{k}^{m}+\frac{i}{2}(m+1-2\alpha), 
  \label{string}
\end{eqnarray}
where $u_{k}^{m}$ are the center of the strings;  
$ \alpha \in \{1,2,\dots, m \}$; 
$ k \in \{1,2,\dots, n_{m} \}$; 
$n_{m}$ is the number of the $m$-strings. 
Assuming (\ref{string}) and taking the product of (\ref{BAE})
 on each string, we obtain 
\begin{eqnarray}
\hspace{-40pt}
&& e\left( \frac{u_{k}^{m}}{m}\right)^{N}
 = (-1)^{(N-n+1)m}
 \prod_{l=1}^{\infty}
 \prod_{j=1}^{n_{l}}
 E_{m l}(u_{k}^{m}-u_{j}^{l}), \label{BAE-pro}
\end{eqnarray}
where 
$e(u):=(u+\frac{i}{2})/(u-\frac{i}{2})$ 
and 
\begin{eqnarray}
E_{m l}(u)&=&e\left(\frac{u}{|l-m|}\right)
e^{2}\left(\frac{u}{|l-m|+2}\right)
\cdots 
e^{2}\left(\frac{u}{l+m-2}\right)
e\left(\frac{u}{l+m}\right)  \nonumber \\  
&& \hspace{-30pt}
\times 
e^{-1}\left(\frac{u}{|l-m|+1}\right)
e^{-1}\left(\frac{u}{|l-m|+3}\right)
\cdots 
e^{-1}\left(\frac{u}{l+m-1}\right) 
\; {\rm if } \; m \ne l, \nonumber \\ 
&=&
e^{2}\left(\frac{u}{2}\right)
e^{2}\left(\frac{u}{4}\right)
\cdots 
e^{2}\left(\frac{u}{2m-2}\right)
e\left(\frac{u}{2m}\right)  \nonumber \\  
&& 
\times 
e^{-1}\left(\frac{u}{1}\right)
e^{-1}\left(\frac{u}{3}\right)
\cdots 
e^{-1}\left(\frac{u}{2m-1}\right) 
\qquad {\rm if } \quad m=l. \label{gonta}
\end{eqnarray}
Taking the logarithm of (\ref{BAE-pro}),  we obtain 
\begin{eqnarray}
\hspace{-40pt}
&& N\theta\left( \frac{u_{k}^{m}}{m}\right)
 = 
 2\pi I_{k}^{m}
 +\sum_{l=1}^{\infty}
 \sum_{j=1}^{n_{l}}
 \Theta_{m l}(u_{k}^{m}-u_{j}^{l}), 
 \label{kitarou}
\end{eqnarray}
where $\theta(u)=2 \arctan (2 u )  $, 
$I_{k}^{m}\in \frac{1}{2}{\mathbb Z}$ and 
\begin{eqnarray}
\Theta_{m l}(u)&=&\theta\left(\frac{u}{|l-m|}\right)
+2\theta\left(\frac{u}{|l-m|+2}\right)
+\cdots 
+2\theta\left(\frac{u}{l+m-2}\right)
 \nonumber \\
&& \hspace{-40pt}
+\theta\left(\frac{u}{l+m}\right)   
- 
\theta\left(\frac{u}{|l-m|+1}\right)
-\theta\left(\frac{u}{|l-m|+3}\right)
-\cdots 
-\theta\left(\frac{u}{l+m-1}\right) \nonumber \\  
&& \hspace{150pt} {\rm if } \quad m \ne l, \nonumber \\
&=&
2\theta\left(\frac{u}{2}\right)
+2\theta\left(\frac{u}{4}\right)
+\cdots 
+2\theta\left(\frac{u}{2m-2}\right)
+\theta\left(\frac{u}{2m}\right)  \nonumber \\  
&& 
- 
\theta\left(\frac{u}{1}\right)
-\theta\left(\frac{u}{3}\right)
-\cdots 
-\theta\left(\frac{u}{2m-1}\right) 
\qquad {\rm if } \quad m=l. \label{kakutan}
\end{eqnarray}
Substituting (\ref{string}) into (\ref{en}), we obtain 
the energy in terms of the centers of the strings: 
\begin{eqnarray}
\hspace{-40pt}
&& E=
J \left\{
\sum_{m=1}^{\infty}
 \sum_{j=1}^{n_{m}}
 \frac{m}{(u_{j}^{m})^{2}+(\frac{m}{2})^{2}}
 -N
 \right\} . 
 \label{enst}
\end{eqnarray} 
Now we shall take the thermodynamic limit 
($N\to \infty$) of (\ref{kitarou}) and (\ref{enst}). 
Let particle, hole and vacancy densities be 
 $\rho_{m}^{p}(u)$, $\rho_{m}^{h}(u)$ 
 and $\rho_{m}(u)$, respectively: 
\begin{eqnarray}
 && \rho_{m}(u):=
     \lim_{N \to \infty}\frac{I_{j+1}^{m}-I_{j}^{m}}
               {N(u_{j+1}^{m}-u_{j}^{m})}
          =\rho_{m}^{h}(u)+\rho_{m}^{p}(u), \nonumber \\ 
&& \rho_{m}^{p}(u):=
  \lim_{N \to \infty}\frac{1}{N(u_{j+1}^{m}-u_{j}^{m})}
.   \label{denc}
\end{eqnarray}
In the thermodynamic limit, (\ref{kitarou}) 
reduces to the following integral equations: 
\begin{eqnarray}
\hspace{-40pt}
&& f_{m}(u)= \rho_{m}^{h}(u)+ 
 \sum_{l=1}^{\infty} 
  A_{l\, m}\rho_{l}^{p}(u) 
 , \label{aint}
\end{eqnarray} 
where $A_{lm}$ is defined as follows 
\begin{eqnarray}
&A_{lm}=A_{ml}=&
[|l-m|]+2[|l-m|+2]+\cdots +2[l+m-2]+[l+m] \nonumber \\ 
&& -[|l-m|+1]-[|l-m|+3]-\cdots -[l+m-1] ,
 \label{taka-operator} 
\end{eqnarray} 
and $[m]$ acts on any function $g(u)$ as follows 
\begin{eqnarray}
 [m]g(u)&=& f_{m}*g(u) 
 \quad  {\rm for} \quad m \ne 0, 
   \nonumber \\ 
\ [0]g(u )&=&g(u ), \label{taka-operator2}
\end{eqnarray} 
where $*$ denotes a convolution 
\begin{eqnarray}
 f_{m}*g(u)&=& \int_{-\infty}^{\infty}
               f_{m}(u-v ) g(v ) d v
 , \label{conta}
\end{eqnarray} 
and $f_{m}(u)$ is defined as follows 
\begin{eqnarray}
 f_{m}(u)=\frac{m}{2\pi \left\{u^{2}+(\frac{m}{2})^{2}\right\}}. 
  \label{katochan}
\end{eqnarray}
The operators (\ref{taka-operator}) and 
 (\ref{taka-operator2}) are $osp(1|2)$ version of 
 the ones introduced by Takahashi \cite{T}. 
 The energy (\ref{enst}) per site becomes 
\begin{eqnarray}
\hspace{-40pt}
&& \frac{E}{N}= J\left\{ \sum_{m=1}^{\infty}
 \int_{-\infty}^{\infty}
 \frac{m}{u^{2}+(\frac{m}{2})^{2}}
 \rho_{m}^{p}(u) 
 d u 
 -1 \right\}. \label{getarou}
\end{eqnarray} 
The entropy per site is given as follows 
\begin{eqnarray}
\hspace{-40pt}
\frac{S}{N} &=&
k_{B}
 \sum_{m=1}^{\infty}
 \int_{-\infty}^{\infty}
 \left\{ 
  (\rho_{m}^{h}(u)+\rho_{m}^{p}(u))
\ln(\rho_{m}^{h}(u)+\rho_{m}^{p}(u)) \right. 
 \nonumber \\ 
&& \left. -\rho_{m}^{h}(u) \ln \rho_{m}^{h}(u)
-\rho_{m}^{p}(u) \ln \rho_{m}^{p}(u)
 \right\} d u, \label{tarou}
\end{eqnarray} 
where $k_B$ is the Boltzmann constant.
In the thermodynamic limit, the free energy $F=E-TS$ 
($T$: temperature) should be minimized with respect to density functions 
(\ref{denc}). Namely, we have
\begin{eqnarray}
\hspace{-30pt} && 0=\frac{ \delta F }{N}=
 J\sum_{m=1}^{\infty} 
  \int_{-\infty}^{\infty} 
    \frac{m}{u^{2}+(\frac{m}{2})^{2}}
    \delta \rho_{m}^{p}(u) d u \label{foint} \\ 
  \hspace{-30pt}  && \hspace{15pt} -
    k_{B} T\sum_{m=1}^{\infty}
     \int_{-\infty}^{\infty} 
     \left\{
    \delta \rho_{m}^{p}(u) 
    \ln \left( 1+\frac{\rho_{m}^{h}(u)}{\rho_{m}^{p}(u)} \right) 
      +
   \delta \rho_{m}^{h}(u) 
   \ln \left( 1+\frac{\rho_{m}^{p}(u)}{\rho_{m}^{h}(u)} \right) 
     \right\} d u . \nonumber
\end{eqnarray}
One can derive the following relation from (\ref{aint}).
\begin{eqnarray}
\delta \rho_{m}^{h}(u)=-\sum_{l=1}^{\infty} 
 A_{ml} \delta \rho_{l}^{p}(u). \label{adel} 
\end{eqnarray}
Combining (\ref{adel}) with (\ref{foint}), we obtain 
the following integral equation 
\begin{eqnarray}
\ln(1+\eta_{m}(u))=
 2\pi \beta J f_{m}(u)+
\sum_{l=1}^{\infty} A_{ml}\ln(1+\eta_{l}(u)^{-1}),\label{sukesan}
\end{eqnarray}
where $\beta=1/k_B T$ and $\eta_{m}(u):=\rho_{m}^{h}(u)/\rho_{m}^{p}(u)$. 

To proceed further, we assume the function $\eta_m(u)$ ($m\ge 1$)
have constant asymptotics in the limits $u \to \infty$.
After performing the Fourier transformation
on both sides of above equation (\ref{sukesan}), we obtain
\begin{equation}
{\mathcal F}[\ln(1+\eta_m)](k)=2\pi J \beta e^{-\frac{m}{2}|k|}
         +\sum_{l=1}^{\infty}{\mathcal F}[A_{m l}](k)
                                {\mathcal F}[\ln(1+\eta_l^{-1})](k),
\label{fourier}
\end{equation}
where $k\in {\mathbb R}$ and
\begin{eqnarray}
{\mathcal F}[A_{ml}](k)&=&\left(1-\frac{1}{2\cosh\frac{k}{2}}\right)
                       B_{ml}(k), \nonumber \\
B_{ml}(k)&=&\left(e^{-\frac{|l-m|}{2}|k|}-e^{-\frac{l+m}{2}|k|}\right)
               \frac{\cosh\frac{|k|}{2}}{\sinh\frac{|k|}{2}}.
\end{eqnarray}
In the above, we adopt
\begin{equation}
{\mathcal F}[f](k)=\int_{-\infty}^{\infty}f(u)e^{-i k u} d u,
\end{equation}
as the Fourier components of a function $f(u)$.
Multiplying both sides in (\ref{fourier}) by
\begin{equation}
B_{nm}^{-1}(k)=\delta_{n,m}-\frac{1}{2\cosh\frac{k}{2}}
                \times
                \left\{
                  \begin{array}{@{\,}ll}
                     \delta_{n,m+1} & \mbox{for $m=1$} \\
                     \delta_{n,m+1}+\delta_{n,m-1} & \mbox{for $m\ge 2$},
                   \end{array}
                \right.    
\label{binverse}
\end{equation}
and taking summation with respect to $m$, we obtain
\begin{eqnarray}
{\mathcal F}[\ln\eta_1](k)&=&
                  \frac{\pi J \beta}{\cosh\frac{k}{2}}-
                  \frac{1}{2\cosh\frac{k}{2}}
                      \left({\mathcal F}[\ln(1+\eta^{-1}_{1})](k)
                               -{\mathcal F}[\ln(1+\eta_{2})](k)\right),
                               \nonumber \\
{\mathcal F}[\ln\eta_m](k)&=&\frac{1}{2\cosh\frac{k}{2}}
                   \left({\mathcal F}[\ln(1+\eta_{m-1})](k)
                               +{\mathcal F}[\ln(1+\eta_{m+1})](k)
                                                \right.  \nonumber \\
                  && \left. -{\mathcal F}[\ln(1+\eta^{-1}_{m})](k)
                     \right) \qquad \qquad \qquad
                       \mbox{for $m\ge 2$}.
\end{eqnarray}
By performing the inverse Fourier
transformation, we obtain the following TBA equation 
\begin{eqnarray}
\ln\eta_1(u)&=& \frac{\pi \beta J}{\cosh\pi u}-
                      K*\ln(1+\eta^{-1}_{1})(u)+
                      K*\ln(1+\eta_{2})(u), \nonumber \\
\ln\eta_m(u)&=&K*\ln(1+\eta_{m-1})(u)-K*\ln(1+\eta_{m}^{-1})(u) 
 \nonumber \\
       &&+K*\ln(1+\eta_{m+1})(u)    \qquad \qquad
          \mbox{for $m\ge 2$},
\label{nlie}
\end{eqnarray}
where the integral kernel $K(u)$ is defined by
\begin{equation}
K(u)=\frac{1}{2\cosh{\pi u}}.
\end{equation}

After performing the Fourier transformation on (\ref{aint}) 
and multiplying by $B_{lm}^{-1}$,  we get
\begin{eqnarray}
{\mathcal F}[\rho_1^p](k)&=&\frac{1}{2\cosh\frac{k}{2}-1}\left(
            1-2\cosh\frac{k}{2}{\mathcal F}[\rho_1^{h}](k)+
                               {\mathcal F}[\rho_2^{h}](k)\right),
                               \nonumber \\
{\mathcal F}[\rho^p_m](k)&=&-\frac{1}{2\cosh\frac{k}{2}-1}\left(
             2\cosh\frac{k}{2}{\mathcal F}[\rho_m^{h}](k)-
                               {\mathcal F}[\rho_{m-1}^{h}](k)
                          -{\mathcal F}[\rho_{m+1}^{h}](k) \right)
                               \nonumber \\
             && \qquad \qquad \qquad \qquad \qquad
                \qquad \qquad \qquad \qquad  \mbox{for } m\ge 2.
\label{inverserho}
\end{eqnarray}
Combining above equations with (\ref{getarou}) and (\ref{tarou}),
and using the fact that
 $\eta_m(u)=\eta_m(-u)$ for $m\ge 1$, 
 we can derive the free  energy per site $f$ as
\begin{equation}
f=J\left(\frac{4\pi}{3\sqrt{3}}-1\right)-k_{B}T\int_{-\infty}^{\infty}
                             R(u)\ln(1+\eta_1(u))du,
\label{free}
\end{equation}
where $R(u)$ is written as
\begin{equation}
R(u)=\frac{2\sinh\frac{4\pi u}{3}}{\sqrt{3}\sinh2\pi u}.
\label{kernel}
\end{equation}
To determine the unique solutions of the TBA equation (\ref{nlie}), 
we need to obtain the asymptotic values 
$\eta_{m}(\pm \infty):=\lim_{u\to\pm\infty}\eta_m(u)$ ($m\ge 1$). 
To determine $\eta_{m}(\pm \infty)$, 
we shall consider the high temperature limit $T\to\infty$.
In this limit, we assume that 
all the functions $\eta_m(u)$ ($m\ge 1$) are 
independent of the parameter $u$. Then (\ref{nlie}) can
be rewritten as follows.
\begin{eqnarray}
&&\eta_1^2=\frac{1+\eta_2}{1+\eta_1^{-1}},  \nonumber \\
&&\eta_{m}^2=\frac{(1+\eta_{m+1})(1+\eta_{m-1})}{1+\eta_m^{-1}}
                                      \qquad \mbox{for $m\ge 2$}.
                                            \label{etahigh}
\end{eqnarray}
The free energy (\ref{free}) must have the following form,
\begin{equation}
 \lim_{T \to \infty} 
 \frac{f}{k_{B} T}=
 -\ln 3,
\end{equation}
which follows from the fact that the freedom of the state per site
is three, namely the entropy per site must be $S=k_{B}\ln 3$ in 
the high temperature limit.
Thus we have $\eta_1=2$ and the
following solutions of (\ref{etahigh}):  
\begin{equation}
\eta_{m}=\frac{m(m+3)}{2}\quad \mbox{for } m\ge 1.
\label{asympt}
\end{equation}

Therefore at any finite temperature ($T>0$), 
the asymptotic values of $\eta_{m}(u)$ 
in (\ref{nlie}) have the same values as (\ref{asympt}):
\begin{equation}
\lim_{u\to\pm\infty}\eta_m=\frac{m(m+3)}{2}\quad \mbox{for } m\ge 1,
\label{asympt2}
\end{equation}
since $\eta_{m}(\pm \infty)$ also satisfy the same 
equations as (\ref{etahigh}) and 
assumed to be independent of $T$.
Consequently, the free energy per site at  any
finite temperatures can be uniquely determined by (\ref{free})
through the TBA equation (\ref{nlie})
and the asymptotic values (\ref{asympt2}).
The above representation for the TBA equation (\ref{nlie}),
the asymptotic values (\ref{asympt2}) and  
the free energy (\ref{free}) can also be confirmed by the 
QTM method, 
which is free from the string hypothesis \cite{IP}. 
One might suspect that our TBA equation (\ref{nlie}) is nothing but 
a certain specialization of the one \cite{MNTT} for the 
Izergin-Korepin model since the affine version of $osp(1|2)$, 
i.e. $osp(1|2)^{(1)}$ has close resemblance to the affine 
Lie algebra $A_{2}^{(2)}$. However, to the author's knowledge, 
so far the TBA equation for the Izergin-Korepin model has been 
constructed \cite{MNTT} so that it reduces to the one for a 
$su(3)$-invariant model in the isotropic limit. 
Therefore, we expect that our TBA equation (\ref{nlie}) is new. 

{}From the above expression, one can analyze the
the low temperature limit of the free energy.
Let  $\varepsilon_m(u)=k_{B}T\ln\eta_{m}(u)$.
In the case $J>0$, we can see that the functions
$\varepsilon_m$ ($m\ge1)$ are always positive by 
applying the iteration method to the TBA equation
(\ref{nlie}).
Therefore from (\ref{sukesan}), 
the functions $\varepsilon_m(u)$ are written,
 in the limit $T \to 0$, as
\begin{equation}
\varepsilon_m(u)=2\pi J f_{m}(u) \qquad \mbox{for } m\ge 1.
\end{equation}
Thus we have
\begin{equation}
\eta_m=\infty, \qquad \rho_m^p=0.
\label{asympt3}
\end{equation}
These results mean that all the spins are in the same direction,
which agrees that the $J>0$ case corresponds to the ferromagnetic regime.
{}From above Eq.~(\ref{asympt3}) and Eq.~(\ref{getarou}), 
we have the ground 
state energy per site $e_0$:
\begin{equation}
e_0=-J.
\end{equation}

While in the case $J<0$,
one sees from the TBA equation (\ref{nlie}) that
$\eta_m\le m(m+3)/2$ ($m\ge 1$). 
Therefore the second term in the equation (\ref{free})
will vanish in the limit  $T\to 0$.
Thus the ground state energy per site $e_0$ is given by
\begin{equation}
e_0=J\left(\frac{4\pi}{3\sqrt{3}}-1\right),
\end{equation}
which coincides with the result in \cite{Mar95-2}.

\section*{Acknowledgments}
\noindent
The authors would like to thank Professor A. Kuniba for encouragement.
This work is supported in part by a Grant-in-Aid for 
JSPS Research Fellows from the Ministry of Education, 
Science, Culture and Sports of Japan. 
%
              

\end{document}